\documentclass[twocolumn,showpacs,floatfix,superscriptaddress,pre]{revtex4}
\usepackage{graphicx}
\usepackage{dcolumn}
\usepackage{bm}

\begin{document}
\title{Comment on ``Equilibrum spin currents: Non-Abelian Gauge Invariance and Color Diamagnetism in Condensed Matter"}

\author{Alexander L\'opez}
\affiliation{Centro de F\'\i sica, Instituto Venezolano de Investigaciones Cient\'ificas, Apartado 21874, Caracas 1020-A, Venezuela}
\author{Ernesto Medina}
\affiliation{Centro de F\'\i sica, Instituto Venezolano de Investigaciones Cient\'ificas, Apartado 21874, Caracas 1020-A, Venezuela}
\affiliation{Departamento de F\'isica, Universidad Central de Venezuela, Caracas, Venezuela}
\author{Nelson Bol\'ivar}
\affiliation{Departamento de F\'isica, Universidad Central de Venezuela, Caracas, Venezuela}
\author{Bertrand Berche}
\affiliation{Laboratoire de Physique des Mat\'eriaux, Universit\'e Henri Poincar\'e, Nancy 1, 54506
Vand\oe uvre les Nancy Cedex, France}


\maketitle

Recently there has been a lot of interest in expressing the Spin-orbit Hamiltonians in terms of a Yang-Mills type formulation, where the corresponding vector potential is non-Abelian. This formulation is very revealing, since the consistent gauge structure of the theory becomes obvious. As the Yang-Mills gauge theory is by now well understood and is the underpinning of well established physical theory, enormous insight can be brought upon new problems. In \cite{Tokatly}, Tokatly drew upon work of $\rm Fr\ddot{o}hlich$ and Studer\cite{Frohlich} to write an explicitly gauge invariant Hamiltonian corresponding to a many body theory with first spin dependent relativistic corrections, giving rise to Spin Orbit (SO) interactions. An explicit form for the non-Abelian gauge potential makes a connection to 2D models of semiconductors  with linear in $k$-vector SO interactions.

In this comment we draw attention to the fact that both the Pauli equation and the gauge formulations of the Rashba and 2D Dresselhaus Hamiltonian cannot be written correctly in a generally gauge invariant fashion\cite{Leurs,MedinaLopezBerche}. We argue that this makes a very strong physical statement: a gauge invariant Hamiltonian would not allow a locally gauge invariant spin current or spin polarization. We show that, in fact, the correct formulation contains and demands a gauge symmetry breaking term that allows for locally gauge invariant matter spin current. The same term is also related to the existence of a gauge potential  as a physical field in the same way as in superconductivity, required by the very identification made in \cite{Tokatly} where it is directly related to a physical $U(1)$ electric field.

The Lagrangian generating the correct Schrodinger equation is 
\begin{eqnarray}
&&{\mathcal L}= \frac{i\hbar}{2}\left (\Psi^{\dagger}\dot{\Psi}-\dot{\Psi}^{\dagger}\Psi\right )\nonumber\\
&-&\Psi^{\dagger}\left (\frac{-g^2}{8m_0}{W}_i^b{W}_i^b+g{W}_0^a\tau^a+e{A}_0\right)\Psi\nonumber\\
&-&\frac{\hbar^2}{2m_0}\left [ {\mathcal D}_j\Psi\right ]^{\dagger}{\mathcal D}_j\Psi
-\frac{e^2}{4m_0}F_{\mu\nu}F_{\mu\nu}-\frac{g^2}{4m_0}{G}^a_{\mu\nu}{ G}^{a}_{\mu\nu},\nonumber
\end{eqnarray}
where $\Psi$ is a Pauli spinor, ${G}_{\mu\nu}^a =\partial_\mu{W}_\nu^a-\partial_\nu{W}_\mu^a
 -\epsilon^{abc}{W}_\nu^b{W}_\mu^c$ and ${F}_{\mu\nu} =\partial_\mu A_\nu-\partial_\nu A_\mu$ are the $SU(2)$ and $U(1)$ field tensors respectively. The new term $\frac{-g^2}{8m_0}{W}_i^b{W}_i^b$ arises when building the covariant derivatives ${\mathcal D}_i=\partial_i-\frac{ie}{\hbar}A_i-\frac{i g}{\hbar}{W}^a_i\tau^a$ that define the gauge potential. For the Pauli Hamiltonian\cite{Tokatly,MedinaLopezBerche}, $g{W}_i^a\tau^a=-(e\hbar/2 m_0c^2)\varepsilon_{iaj}{ E}_j\tau^a$,  where the $\tau^a$ are the symmetry generators for $SU(2)$. This Lagrangian is not gauge invariant, as can be seen from the presence of the quadratic in gauge field term. This term restricts the gauge invariance to a smaller set of transformations. In the same spirit of the Chern-Simons theory, it can be related to the gauge invariant form of the action, by rewriting the extra terms as a choice of gauge plus a surface term. The gauge choice amounts to fixing the norm of the spinor wavefunction plus the condition $\partial_iW_i^b=0$, an $SU(2)$ version of the Coulomb gauge. Such gauge is also required from Maxwell's equations\cite{Leurs}. One can now derive the conserved current $\mathcal J^a_{i}$ in the ordinary sense, from the equations of motion. This is the full spin current carried both by matter and radiation i.e. ${\cal J}\!\!\!\!\!{\mathcal J}^a={\boldsymbol J}^a_M+{\boldsymbol J}^a_R$. The spatial (spin current) components of the current density then follow as
${\mathcal J}_i^a=\Psi^{\dagger} \left ( \frac{g^2}{4m_0}{W}^a_i\right )\Psi+\frac{g^2}{m_0}\varepsilon^{abc}{W}^b_\nu{G}^{c}_{\nu i}
-\frac{i\hbar g}{2m_0}\left [ \left (\tau^a\Psi\right)^{\dagger}{\mathcal D}_i\Psi
-\left ({\mathcal D}_i\Psi\right )^{\dagger}\left (\tau^a\Psi\right ) \right ]$,
and the spin polarization (time component)
${\cal J}^a_0=\Psi^{\dagger}g\tau^a\Psi+\frac{g^2}{m_0}\varepsilon^{abc}{W}^b_j{G}^{c}_{j0}$.
Three terms can be distinguished in the spin current; the third term has the canonical form expected for the material current. The second term is the radiative contribution and the first term comes from the gauge symmetry breaking contribution. The equilibrium currents, discussed in references \cite{Tokatly,RashbaEquil,MedinaLopezBerche}, arise in relation to the radiative contribution, cubic in the non-Abelian potential. 

In reference\cite{Tokatly}, Tokatly argues on the fact that the variational definition of the current density leaves no room for any ambiguity on the definition of the spin current density. We would like to stress that this statement is only valid in the formulation stated here  since we have a gauge symmetry broken (or fixed gauge) formulation. Otherwise, gauge transformations would change the matter (spin) and radiation (torque) content i.e. the matter currents are gauge covariant. A similar gauge symmetry scenario to the one presented here is materialized in superfluid condensates of neutral Bosons in the Helium 3 B phase, where one encounters a $SU(2)$ Anderson-Higgs mechanism\cite{Frohlich}.

\end{document}